\documentclass[twocolumn,apj]{emulateapj}
\usepackage{natbib}
\usepackage{amsmath,amssymb}
\usepackage{txfonts}
\usepackage{color,hyperref}

\hypersetup{colorlinks,breaklinks,
            linkcolor=blue,urlcolor=blue,
            anchorcolor=blue,citecolor=blue}

\newcommand{\mrm}[1]{\mathrm{#1}}

\newcommand{\sinc}{\mathrm{sinc}}

\newcommand{\lmax}{\ensuremath{\ell_\mathrm{max}}}

\newcommand*\bra[1]{\left(#1 \right)}
\newcommand{\bea}{\begin{eqnarray}}
\newcommand{\eea}{\end{eqnarray}}
\newcommand{\bean}{\begin{eqnarray*}}
\newcommand{\eean}{\end{eqnarray*}}

  {}

\received{2017 March 8}
\revised{2017 June 26}
\accepted{2017 June 28}

\shorttitle{The galaxy count correlation function in redshift space revisited}
\shortauthors{Campagne, Plaszczynski, \& Neveu}

\begin{document}

\title{The galaxy count correlation function in redshift space revisited}





\author{J.-E. Campagne\footnote{campagne@lal.in2p3.fr}, S. Plaszczynski and J. Neveu}
\affil{LAL, Univ. Paris-Sud, CNRS/IN2P3, Universit\'e Paris-Saclay, F-91898 Orsay, France}

\begin{abstract}
In the near future, cosmology will enter the wide and deep galaxy survey era, enabling high-precision studies of the large-scale structure of the universe in three dimensions. To test cosmological models and determine their parameters accurately, it is necessary to use data with exact theoretical expectations expressed in the observational parameter space (angles and redshift). The data-driven, galaxy number count fluctuations on redshift shells can be used to build correlation functions  $\xi(\theta, z_1, z_2)$  on and between shells to probe the baryonic acoustic oscillations and distance-redshift distortions, as well as gravitational lensing and other relativistic effects. To obtain a numerical estimation of $\xi(\theta, z_1, z_2)$ from a cosmological model, it is typical to use either a closed form derived from a tripolar spherical expansion or to compute the power spectrum $C_\ell(z_1,z_2)$ and perform a Legendre polynomial $P_\ell(\cos\theta)$ expansion. 
Here, we present here a new derivation of a $\xi(\theta, z_1, z_2)$ closed form 
using the spherical harmonic expansion and proceeding to an infinite sum over multipoles thanks to an addition theorem. We demonstrate that this new expression is perfectly compatible with the existing closed forms but is simpler to establish and manipulate. We provide formulas for the leading density and redshift-space contributions, but also show how Doppler-like and lensing terms can be easily included in this formalism.
We have implemented and made publicly available a software for computing   those correlations efficiently, without any Limber approximation,  and validated this software with the  \texttt{CLASSgal} code. It is available at \url{https://gitlab.in2p3.fr/campagne/AngPow}.

\end{abstract}

\section{Introduction}
In the near future, wide and deep surveys of galaxies, performed with, for instance,
the Dark Energy Spectroscopic Instrument (DESI)
\citep{2013arXiv1308.0847L},  the Large Synoptic Survey Telescope
(LSST) \citep{2008arXiv0805.2366I}, and the \textit{Euclid} satellite
\citep{2011arXiv1110.3193L}, will map the galaxy density field with
unprecedented precision. They will produce catalogs containing
information on the large-scale structure of the universe in three
dimensions: the angular position $\mathbf{n}$ and the redshift $z$ of
each galaxy. Most of the cosmological information in clustering studies is contained within two-point
functions of matter overdensities, o,r equivalently, their Fourier power spectra. 
Data are most easily analyzed and understood in the real space, but in
linear perturbation theory, models are expressed in the Fourier space
 (where perturbation modes evolve independently). 

When observing galaxies, one does not really have a direct measurement of the three
$(x,y,z)$ Euclidean coordinates: redshifts must be translated into
positions, which requires a fiducial cosmological model that introduces theory within observations.
Furthermore, the model neglects the peculiar velocities of galaxies and
produces the so-called "redshift-space distortions" (RSDs),  which  have been better expressed in
the Fourier space, following the pioneering work of \citet{1987MNRAS.227....1K}.

On the theory side, thanks to the work of \citet{2008cmb..book.....D,2009PhRvD..80h3514Y,2010PhRvD..82h3508Y,2011PhRvD..84d3516C,2011PhRvD..84f3505B},  a complete gauge-invariant formalism exists that includes all the linear contributions affecting overdensities measured at a given redshift. This formalism allows us to compute the power spectra on and between spherical
"shells"  at a given redshift possibly smeared by some selection function.

The use of redshift-dependent angular correlation functions
on and between two redshift shells as an ensemble average,
\begin{equation}
  \label{eq:ctdef}
\xi(\theta, z_1, z_2)=\langle
\Delta(\mathbf{n}_1,z_1) \Delta(\mathbf{n}_2,z_2)
\rangle \quad \mathrm{with}\quad  {\mathbf{n}_1.\mathbf{n}_2 = \cos\theta}
  \end{equation}
then looks  appealing both for theoretical and experimental reasons. 
More precisely, Equation (\ref{eq:ctdef}) is computed using the galaxy
number count perturbation in a 3D spherical volume
$\Delta(\mathbf{n},z)= (N(\mathbf{n},z)-\langle N\rangle(z))/\langle
N\rangle(z)$ in two directions ($\textbf{n}_1, \textbf{n}_2$) at two
different redshifts ($z_1, z_2$) under the constraint
$\mathbf{n}_1.\mathbf{n}_2 = \cos\theta$.

To get a compact expression of the $\xi(\theta, z_1, z_2)$ function that is numerically tractable, one can express $\Delta(\mathbf{n},z)$ in the Fourier space, and use the $\xi$ symmetry to perform a tripolar spherical harmonics expansion, which is simplified in specific coordinate systems. This is the path followed, for instance, in \citet{1998ApJ...498L...1S,2004ApJ...614...51S,2008MNRAS.389..292P,2012PhRvD..86f3503M,2012JCAP...10..025B}. This approach has also been  followed in the context of two-point correlation functions between two different populations of galaxies \citep[e.g.][]{2014PhRvD..89h3535B, 2014CQGra..31w4002B}.

Alternatively, to obtain numerical values of $\xi(\theta, z_1, z_2)$, one may also proceed first by expanding $\Delta(\mathbf{n},z)$ at a fixed redshift $z$ on a spherical harmonic basis to determine an 
angular power spectrum $C_\ell(z_1, z_2)$ between two shells at $z_1$ and $z_2$, and secondly by computing
\begin{equation}
\xi(\theta, z_1, z_2) = \frac{1}{4\pi}  \sum_{\ell=0}^\infty (2\ell +1)  P_\ell(\cos\theta) C_\ell(z_1,z_2),
\label{eq-Ctheta-z1z2}
\end{equation}
with $P_\ell(x)$ being the $\ell$th Legendre polynomial. This is the path followed, for instance, in references \citet{ClassGal, 2014JCAP...01..042D, 2015JCAP...10..070M, 2017JCAP...02..020L}. In practice, the sum in Equation (\ref{eq-Ctheta-z1z2}) must be truncated at some $\ell=\lmax$
value, which leads to spurious oscillations and requires  some classically smooth
windowing ("apodization").

Looking at the two methods, it is legitimate to ask whether one can obtain a compact expression of $\xi(\theta,  z_1, z_2)$  using Equation (\ref{eq-Ctheta-z1z2}) and bypassing the intermediate computation of $C_\ell$. It is the purpose of this article to show that
such a direct method can be set up efficiently as a point-to-point equivalent to the previously existing tripolar spherical expansion, and be suitable for numerical implementation. From this perspective, we have released the \verb|Angpow| code \citep{2017arXiv170103592C} to proceed with some numerical tests and a comparison with the publicly available code \verb|CLASSgal| \citep{ClassGal}. 

In the following sections, after a brief review of the computation of $\xi(\theta, z_1, z_2)$, with RSD included by tripolar spherical expansion, we proceed to the derivation of the power spectrum $C_\ell(z_1,z_2)$  and then expose in detail our new procedure of expansion. We show that both compact expressions agree point-to-point and discuss the advantage of the new expression. Some extensions that include other effects such as redshift selection functions and physics processes like sub-leading RSD (or with similar expression) and matter lensing, are also discussed.  Then,  the new computation is validated within the framework of the \texttt{Angpow} code against the \texttt{CLASSgal} code using RSD contribution and Gaussian redshift selection functions. We show some results of the $\xi(\theta,  z_1, z_2)$ computation in the measurement space ($\theta$, $z_2-z_1$).
\section{Galaxy count correlation function}
The two-point correlation function is defined as an ensemble average 
\begin{equation}
\xi(\mathbf{n}_1, \mathbf{n}_2,z_1,z_2)=\langle
\Delta(\mathbf{n}_1,z_1) \Delta(\mathbf{n}_2,z_2)
\rangle
\label{eq:corfuncdef}
\end{equation}
with the constraint $\mathbf{n}_1.\mathbf{n}_2 = \cos\theta$ as we assume a homogeneous and isotropic background.
$\Delta(\mathbf{n},z) $ is the galaxy fractional number overdensity in an elementary volume pointed by the observer  in the direction $\mathbf{n}$ at the redshift $z$ (using $\mathbf{r}= r(z)\ \mathbf{n}$). 
\subsection{The relativistic linear theory}
Linear approximation perturbation theory \citep{2008cmb..book.....D,2009PhRvD..80h3514Y,2010PhRvD..82h3508Y,2011PhRvD..84d3516C,2011PhRvD..84f3505B} gives the different contributions to $\Delta(\mathbf{n},z) $, among them the density fluctuations and the redshift RSD that we consider here for simplicity (see Section \ref{sec:extansions} for the introduction of other effects). It yields  
\begin{equation}
\Delta(\mathbf{n},z)= D(\mathbf{n},z) - \frac{1}{{\cal H}(z)} \partial_r (\mathbf{V}(\mathbf{n},z).\mathbf{n}).
\label{eq-DeltaZ}
\end{equation}

The Fourier transform of this equation reads ($\mathbf{k} = k\ \hat{\mathbf{k}}$) 
\begin{equation}
\Delta(\mathbf{k}, z) = D(\mathbf{k}, z) - \frac{k}{{\cal H}(z)}\ (\hat{\mathbf{k}}.\mathbf{n})^2 \  V(\mathbf{k}, z).
\label{eq-Delta-fourier-space}
\end{equation}
with  $D(\mathbf{k}, z)$ as the density fluctuation contribution in the comoving gauge  and  $V(\mathbf{k},z)$ as the velocity potential  in the longitudinal gauge  such that  $\mathbf{V}(\mathbf{k})= -i\hat{\mathbf{k}}V(\mathbf{k})$,  and ${\cal H}(z) = a(z) H(z)$ is the comoving Hubble parameter. To alleviate the notations, we have only shown the redshift $z$, while the $ z$ dependence is given through the conformal time $\tau(z)$ \citep{2011PhRvD..84f3505B,ClassGal}, defined as $\tau(z) = \tau_0 - r(z)$, with $r(z)$ being the radial comoving distance at redshift $z$ and $\tau_0$ being the conformal age of the universe. 

The $D(\mathbf{k}, z)$ and $V(\mathbf{k}, z)$ perturbations are related by transfer functions to some  random metric perturbation that we take to be the initial Bardeen potential $\Psi_\mathrm{in}$  such that
 
 \begin{eqnarray}
D(\mathbf{k}, z) &= T_D(k,z) \Psi_{in}(\mathbf{k}) \nonumber \\
D(\mathbf{k}, z) &= T_V(k,z)  \Psi_{in}(\mathbf{k})
\label{eq-DV-transfer}
\end{eqnarray}
 
The continuity equation gives the following relation between the matter density and the velocity transfer functions:
 
\begin{equation}
T_V(k,z) = - \frac{ {\cal H}(z)}{k} f_a(z) T_D(k,z)
\end{equation}
 
with 
$$
f_a(z) \equiv d \log G(a(z))/d \log a(z) \qquad a(z)=1/(1+z).
$$
For simplicity, we have set the linear redshift-dependant bias $b(z)$ to 1. Equation (\ref{eq-Delta-fourier-space}) yields
 
\begin{equation}
\Delta(\mathbf{k},z) = \Psi_{in}(\mathbf{k}) T_D(k,z) \left(  1 + f_a(z) (\hat{\mathbf{k}}.\mathbf{n})^2 \right).
\end{equation}
 
One defines the primordial power spectrum $P_{\mathrm{in}}(k)$ according to the statistical property of the $\Psi_{\mathrm{in}}(\mathbf{k})$ field \citep{2008cmb..book.....D,2011PhRvD..84f3505B}
\begin{equation}
\langle \Psi_{\mathrm{in}}(\mathbf{k}) \Psi_{\mathrm{in}}(\mathbf{k}^\prime)\rangle = P_{\mathrm{in}}(k) \delta(\mathbf{k}+\mathbf{k}^\prime)
\end{equation}
Finally, we introduce the approximation valid at low redshift but compatible with the next generation of galaxy surveys, using the growth factor $G(z)$ and the power spectrum $P|_{z=0}(k)$ at $z=0$
 
\begin{equation}
P|_{z=0}(k) G(z_1) G(z_2) \approx P_{\mathrm{in}}(k)   T_D(k,z_1)   T_D(k,z_2).
\end{equation}

We recall in the next section how a closed form of the two-point
correlation function using tripolar spherical expansion has been derived.
\subsection{Tripolar spherical harmonics expansion}
\label{sec:tripolar}
The correlation function is the Fourier transform of the power spectrum:
\begin{equation}
\begin{split}
\xi(\mathbf{n}_1, \mathbf{n}_2,z_1,z_2) &=  G(z_1) G(z_2)  \int \frac{\mathtt{d}\mathbf{k}}{(2\pi)^{2/3}}\ e^{-i \mathbf{k}.(\mathbf{r}_2-\mathbf{r}_1)} P|_{z=0}(k)
 \\
&\times \left\{
1 + f_a(z_1) (\hat{\textbf{k}}.\mathbf{n}_1)^2
\right\} 
\left\{
1 + f_a(z_2) (\hat{\textbf{k}}.\mathbf{n}_2)^2 
\right\}.
\end{split}
\end{equation}
Using the Legendre polynomial  expansion of $x^2$, i.e. $x^2=1/3 + 2/3 P_2(x)$, one finds the following expression ($\mathbf{R} \equiv \mathbf{r}_2 - \mathbf{r}_1 \equiv R\ \mathbf{n_{12}}$):
\begin{equation}
\begin{split}
\xi(\mathbf{n}_1, \mathbf{n}_2,z_1,z_2) & =  G(z_1) G(z_2) \int \frac{\mathtt{d}\mathbf{k}}{(2\pi)^{2/3}}\ e^{-i\ (kR)\  \hat{\mathbf{k}}.\mathbf{R}} P|_{z=0}(k)
 \\
&\times \left\{
1 + \frac{f_a(z_1)}{3} +\frac{2f_a(z_1)}{3} P_2(\hat{\textbf{k}}.\mathbf{n}_1)
\right\}\\
&\times  
\left\{
1 + \frac{f_a(z_2)}{3} +\frac{2f_a(z_2)}{3} P_2(\hat{\textbf{k}}.\mathbf{n}_2)
\right\}.
\end{split}
\end{equation}

\begin{figure}\centering
\includegraphics[width=0.7\linewidth]{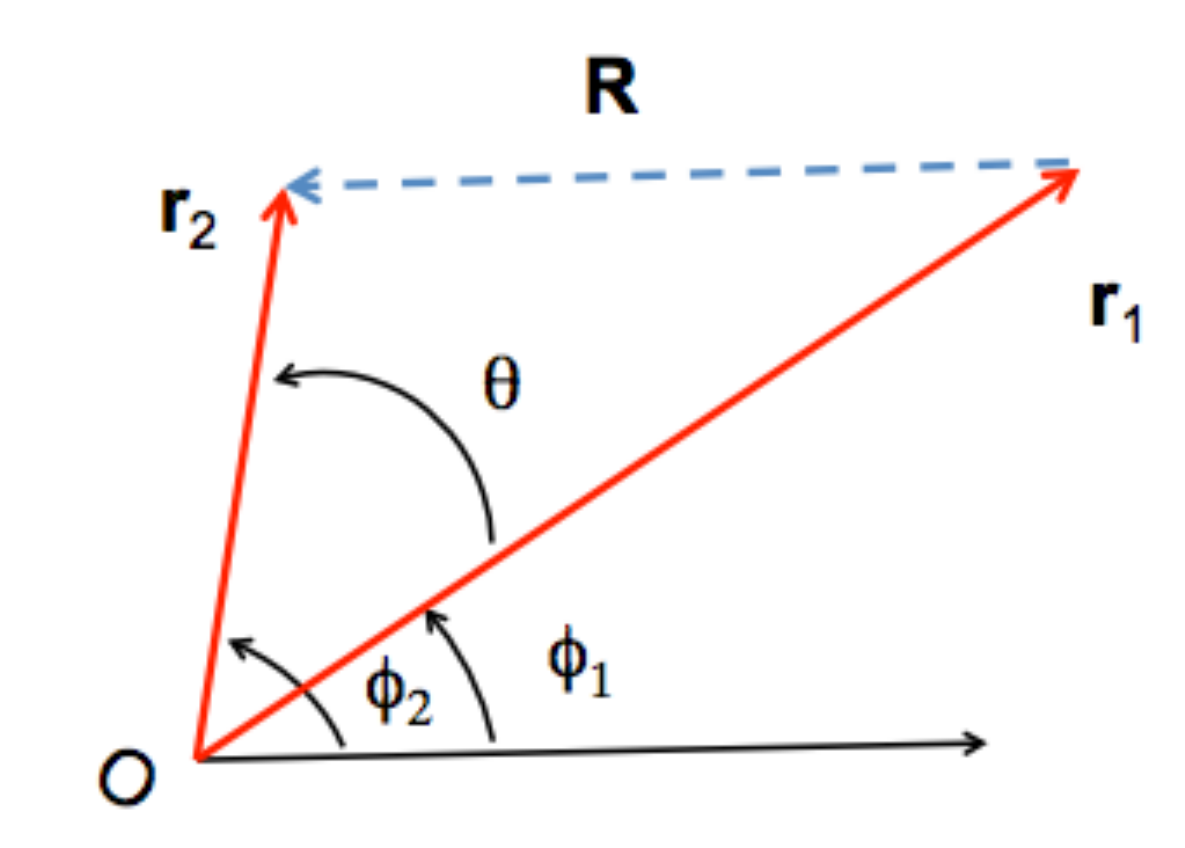}
\caption{Definition of the coordinate system used by observer O to point in two directions,  $\mathbf{r}_1$ and $\mathbf{r}_2$, which form a plane. Here we assume $\phi_1\leq \phi_2$. Implicitly $\phi_1$ ($\phi_2$) depends on $r_1$ $(r_2)$, $R$ and $\theta=\phi_2-\phi_1$. $\mathbf{R}$ is parallel to the reference axis of the $\phi$ angles.
}
\label{fig-angles}
\end{figure}

Closed forms were established, e.g., in reference
\citet{1998ApJ...498L...1S} when $z_1, z_2 \ll 1$ and without this
restriction, as well as and more generally in the context of wide-angle surveys \citep{2000ApJ...535L...1M,2004ApJ...615..573M,2004ApJ...614...51S, 2008MNRAS.389..292P,
  2010MNRAS.409.1525R,2012JCAP...10..025B,2012PhRvD..86f3503M}.
Exploiting the invariance of $\xi$ with respect to the rotation of the
triangle $(\mathbf{r}_1, \mathbf{r_2}, \mathbf{R})$, one can get an expansion on the orthogonal basis of the tripolar spherical harmonic functions defined as \citep{Varshalovich:1988ye, 2004ApJ...614...51S, 2008MNRAS.389..292P,2010MNRAS.409.1525R}

\bea
S_{\ell_1\ell_2\ell}(\mathbf{n}_1,\mathbf{n}_2,\mathbf{n}_{12}) &=& \sum_{m_1,m_2,m} 
\begin{pmatrix}
\ell_1 & \ell_2& \ell \\
m_1 & m_2 & m
\end{pmatrix} \nonumber \\
&\times& C_{\ell_1m_1}(\mathbf{n}_1)C_{\ell_2m_2}(\mathbf{n}_2)C_{\ell m}(\mathbf{n}_{12}),
\label{eq:S_l1l2l}
\eea

with the Wigner 3-j symbol in parentheses and 

$$
C_{\ell m}(\mathbf{n}_{12}) = \sqrt{\frac{4 \pi}{2\ell +1}} Y_{\ell m}(\mathbf{n}_{12}).
$$

Using a specific coordinate system (Figure~\ref{fig-angles}), the tripolar spherical expansion reduces to a compact form \citep{2004ApJ...614...51S, 2008MNRAS.389..292P, 2012PhRvD..86f3503M}
\begin{equation}
\begin{split}
\xi(z_1,z_2,\phi_1,\phi_2) = & G(z_1)G(z_2)\\
\times  \sum_{m_1m_2=\ 0,1,2}&\left[ a_{m_1m_2} \cos(m_1\phi_1)\cos(m_2\phi_2)\right.\\
& + \left. b_{m_1m_2} \sin(m_1\phi_1)\sin(m_2\phi_2) \right]\\
\end{split}
\label{eq:xi-tripolar-expansion}
\end{equation}
The $a_{m_1m_2}$ and $b_{m_1 m_2}$ coefficients (see Appendix \ref{sec:ab-coeff}) depend on
\begin{enumerate}
\item $f_a(z_1)$, $f_a(z_2)$ and
\item  $\zeta_\ell^m(R)$, integral functions defined as 
\begin{equation}
\zeta_\ell^m(R)=\int \frac{\mathtt{d}k}{2\pi^2} k^m j_\ell(k R) P_{z=0}(k)
\label{eq:zeta-func}
\end{equation}
with $j_\ell(x)$ as the spherical  Bessel function of order $\ell$, and
\begin{equation}
R(r_1,r_2,\theta) = \sqrt{r_1^2+r_2^2 - 2 r_1 r_2 \cos\theta}.
\label{eq:R-func}
\end{equation}
In the case of matter density and RSD contributions one needs to compute $\zeta_m^2(R)$ with $m=0,2,4$, exhibiting an integrand proportional to  $k^2\ P_{z=0}(k)$.
\end{enumerate}
One can define $\phi_1$ and $\phi_2$ as functions of $r_1$, $r_2$ and $\theta$ as
\begin{eqnarray}
\phi_1 &=& \arcsin \left( \frac{r_2}{R} \sin\theta\right)  \nonumber \\
\phi_2 &=& \phi_1 + \theta = \arcsin \left( \frac{r_1}{R} \sin\theta\right) 
\end{eqnarray}
so that it is justified to consider $\xi$ as function of $(z_1, z_2, \theta)$. 

\section{Spherical harmonic space}
\label{sec:direct-method}

We present now a new derivation of a closed form of the two-point correlation function using the spherical harmonic expansion of the galaxy count.

\subsection{The power spectrum $C_\ell(z_1,z_2)$}
At fixed redshift value $z$, the function $\Delta(\mathbf{n},z)$
(\ref{eq-DeltaZ}) can be expanded using spherical harmonic functions as  
\begin{equation}
\Delta(\mathbf{n},z) = \sum_{\ell=0}^\infty a_{\ell m}(z) Y_{\ell m}(\mathbf{n})
\label{eq-alm-general}
\end{equation}
The $a_{\ell m}(z)$ coefficients are related to the Fourier transform of $\Delta(\mathbf{n},z)$ by the following relation:
\begin{equation}
a_{\ell m}(z) = i^\ell \sqrt{\frac{2}{\pi}} \int d\mathbf{k} \  \Delta(\mathbf{k},z) j_\ell(k r(z)) Y^\ast_{\ell m}(\hat{\mathbf{k}}).
\end{equation}
Using the identity \citep{2009PhRvD..80h3514Y}

\begin{equation}
i\,(\hat{\mathbf{k}}.\mathbf{n})\ e^{i\ kr\ (\hat{\mathbf{k}}.\mathbf{n})} = \frac{\partial}{\partial(kr)} e^{i\ kr\ (\hat{\mathbf{k}}.\mathbf{n})}  \;,
\label{eq:kn2dkr}
\end{equation} 

then the spherical harmonic contributions from the density fluctuation  and the velocity gradient read
\begin{eqnarray}
a_{\ell m}^{D}(z) &=&  i^\ell  \sqrt{\frac{2}{\pi}}\int \mathrm{d}\mathbf{k}\  D(\mathbf{k}, z) j_\ell(k r(z)) Y^\ast_{\ell m}(\hat{\mathbf{k}}) \nonumber \\
a_{\ell m}^{\partial V}(z) &=& i^\ell \sqrt{\frac{2}{\pi}}  \int \mathrm{d}\mathbf{k}\ k V(\mathbf{k}, z) j_\ell^{\prime\prime}(k r(z)) Y^\ast_{\ell m}(\hat{\mathbf{k}}).
\end{eqnarray}
The $j_\ell^{\prime\prime}(x)$ is the second derivative of the $j_\ell(x)$ function. The spherical harmonics expansion of Equation  (\ref{eq-DeltaZ}) reads
 
\begin{equation}
\begin{split}
a_{\ell m}(z) =i^\ell \sqrt{\frac{2}{\pi}}  & \int d\mathbf{k} \Psi_{in}(\mathbf{k}) T_D(k,z)  \\
& \times \left[ 
 j_\ell(k r(z)) - f_a(z) j_\ell^{\prime \prime}(k r(z))
\right] Y^\ast_{\ell m}(\hat{\mathbf{k}}).
\end{split}
\label{eq-almRSD}
\end{equation}
 
The angular power spectrum $C_\ell(z_1,z_2)$ is obtained as an
ensemble average of the correlation between two $a_{\ell m}(z)$
\begin{equation}
\begin{split}
C_\ell(z_1,z_2) &= \langle a_{\ell m}(z_1) a_{\ell m}^\ast(z_2)\rangle\\
&= G(z_1) G(z_2) \frac{2}{\pi} \int \mathrm{d}k\ k^2\  P|_{z=0}(k)  \\
 & \times \left\{
   j_\ell(k r_1) - f_a(z_1) j_\ell^{\prime\prime}(k r_1)
 \right\} \left\{
   j_\ell(k r_2) - f_a(z_2) j_\ell^{\prime\prime}(k r_2)
 \right\}.
\end{split}
\end{equation}
where $r_i = r(z_i)$. 
%
\subsection{The new $\xi$ expansion}
\label{sec:new-expansion}
Using the statistical isotropy of our universe, the correlation function $\xi$ may be computed by summing up the $C_\ell$'s \citep{2008cmb..book.....D, ClassGal,2014JCAP...01..042D,2017JCAP...02..020L} according to Equation (\ref{eq-Ctheta-z1z2}). The expression of the $\xi$ function reads
\begin{equation}
\begin{split}
\xi(\theta, & z_1, z_2) = G(z_1) G(z_2) \frac{1}{2\pi^2} \int \mathrm{d}k\ k^2 P|_{z=0}(k)  \\ 
&\times \left\{\sum_{\ell=0}^\infty (2\ell+1) P_\ell(\cos\theta) j_\ell(kr_1)j_\ell(kr_2) \right. \\
& - f_a(z_2)\sum_{\ell=0}^\infty  (2\ell+1) P_\ell(\cos\theta) j_\ell(kr_1)j^{\prime\prime}_\ell(kr_2) \\
& - f_a(z_1) \sum_{\ell=0}^\infty  (2\ell+1) P_\ell(\cos\theta) j_\ell(kr_2)j^{\prime\prime}_\ell(kr_1) \\
&\left. + f_a(z_1)f_a(z_2) \sum_{\ell=0}^\infty  (2\ell+1) P_\ell(\cos\theta) j^{\prime\prime}_\ell(kr_1)j^{\prime\prime}_\ell(kr_2) 
\right\}.
\end{split}
\label{eq-Ctheta-ext}
\end{equation}
We define the function $A(x_1, x_2,\theta)$ by the following result \citep[Equation (10.1.45)]{abramowitz+stegun} \footnote{The same relation is also available at the following NIST web page: \href{http://dlmf.nist.gov/10.60}{http://dlmf.nist.gov/10.60}}
\begin{equation}
\begin{split}
A(x_1, x_2, \theta)&= \sum_{\ell=0}^\infty  (2\ell+1) P_\ell(\cos\theta) j_\ell(x_1)j_\ell(x_2) \\
 &= \textrm{sinc}(R(x_1,x_2,\theta))
\end{split}
\label{eq-abra}
\end{equation}
where $\sinc (x)=\sin(x)/x = j_0(x)$ and $x_i=k r_i$.  It yields that the $\xi$ function can be expressed as the following expansion:
\begin{equation}
\begin{split}
\xi(\theta, & z_1, z_2) = G(z_1) G(z_2) \frac{1}{2\pi^2} \int \mathrm{d}k\ k^2\ P|_{z=0}(k)  \\
& \times  \left\{  A(x_1, x_2, \theta)  - f_a(z_2) \frac{\partial^2 A(x_1, x_2, \theta)}{\partial x_2^2}
 -  f_a(z_1)  \frac{\partial^2 A(x_1, x_2,\theta)}{\partial x_1^2} \right. \\
&\left. + f_a(z_1)f_a(z_2)  \frac{\partial^4 A(x_1, x_2,\theta)}{\partial x_1^2\partial x_2^2}
\right\}
\end{split}
\label{eq-Ctheta-Afunc}
\end{equation}
which establishes our main result. The $A$-function and its
derivatives  depend on $k r_1$, $k r_2$ and $\cos \theta$ and on the
derivative of $\sinc(x)$ (Appendix \ref{app-afunc}). 

Equations (\ref{eq:xi-tripolar-expansion}) and (\ref{eq-Ctheta-Afunc}) are two expansions that are in principle of the same $\xi$ function, and indeed in  Appendix \ref{app:equality} we demonstrate formally that it is case.
However, we find that our formulation is more straightforward and we provide a public implementation within the \verb\Angpow\ software, which will be discussed in Section \ref{sec:angpow}.
It also provides a natural framework to incorporate the relativistic terms, as exemplified next. 
\subsection{Beyond the Kaiser RSD contribution to galaxy count}
\label{sec:extansions}
So far, we have only considered the matter density
fluctuation and the main RSD contributions that lead to expression
(\ref{eq-DeltaZ}).
\citet{2008cmb..book.....D}, \citet{2009PhRvD..80h3514Y},
\citet{2010PhRvD..82h3508Y}, \citet{2011PhRvD..84d3516C} and
\citet{2011PhRvD..84f3505B} described in detail other expressions.
It is not the purpose of this section to review all of them but rather
to show with two examples how expression (\ref{eq-Ctheta-Afunc}) can be extended.

\subsubsection{Doppler-like term}
\label{sec:Vn-terms}
The first case concerns terms proportional to
$\mathbf{V}(\mathbf{n},z).\mathbf{n}$, e.g. a sub-dominant RSD (except
at low redshift), a boost term or a selection function as considered, for
instance, in \citet{2008MNRAS.389..292P,2010MNRAS.409.1525R,2012PhRvD..86f3503M, ClassGal, 2016arXiv160203186R}. The perturbation can be described with a generic $\alpha(z)$ function as
\begin{equation}
\Delta_{\mathbf{V.n}}(\mathbf{n},z) = - \frac{\alpha(z)}{r(z)} \mathbf{V}(\mathbf{n},z).\mathbf{n}
\label{eq-Vn-contrib}
\end{equation}
Such a contribution in Fourier space gives rise to a term proportional to $\hat{\mathbf{k}}.\mathbf{n}$ 
\begin{equation}
\Delta_{\mathbf{V.n}}(\mathbf{k}, z) = i \frac{\alpha(z)}{r(z)} (\hat{\mathbf{k}}.\mathbf{n}) V(\mathbf{k}, z)
\end{equation}
and leads to contributions to expansion (\ref{eq:xi-tripolar-expansion})
discussed in \citet{2012PhRvD..86f3503M}. However, using the identity
(\ref{eq:kn2dkr}), one can deduce that in spherical harmonic space it
yields a contribution to Equation (\ref{eq-almRSD}) of the form
 
\begin{equation}
\begin{split}
a_{\ell m}^{\mathbf{V.n}}(z) = - i^\ell \sqrt{\frac{2}{\pi}} & \int d\mathbf{k} \Psi_{in}(\mathbf{k}) T_D(k,z)  Y^\ast_{\ell m}(\hat{\mathbf{k}}) \\
& \times \frac{\alpha(z){\cal H}(z)f_a(z)}{k r(z)} j_\ell^\prime(k r(z))
\end{split}
\label{eq-alm-V}
\end{equation}

which exhibit a $j_\ell^\prime(x)$ factor. Then, correspondingly, this leads to contributions to expression (\ref{eq-Ctheta-Afunc}) due to sums over $\ell$ similar to those shown in Equation (\ref{eq-Ctheta-ext}) which may be expressed as derivatives of the $A$ function (\ref{eq-abra}) as
\begin{equation}
\sum_{\ell=0}^\infty  (2\ell+1) P_\ell(\cos\theta) j_\ell^\prime(x_i)j^{(n)}_\ell(x_j) = \frac{\partial^{n+1}A}{\partial x_i\partial^n x_j}
\label{eq-Doppler-term}
\end{equation}
with $j^{(n)}_\ell(x)$ the $n$-th derivative of $j_\ell(x)$ with respect to $x$. This result can be generalized to any power of $(\hat{\mathbf{k}}.\mathbf{n})^p$ that is converted to a $p$-th derivative of the $A(x_i, x_j,\theta)$ function with respect to  $x_i$ (and $x_j$).
\subsubsection{Lensing term}
The second case concerns integral terms as the solid angle distortion
generated by gravitation lensing (e.g.,
\citealt{2011PhRvD..84f3505B,2014CQGra..31w4002B,2015JCAP...10..070M})
which reads
\begin{equation}
\begin{split}
\Delta_\mathrm{lens}(\mathbf{n},z) = -\frac{1}{r(z)}\int_0^{r(z)} \mathrm{d}r^\prime & \left(\frac{r(z) -r^\prime}{r^\prime} \right)\\
& \times  \Delta_\Omega\left[ (\Psi+\Phi)(r^\prime\mathbf{n},\tau_0-r^\prime) \right]
\end{split}
\label{eq-Delta-lens}
\end{equation}
with $\Psi$ and $\Phi$ as the Bardeen fields \citep{1980PhRvD..22.1882B}
corresponding to the temporal and spatial metric perturbations in the longitudinal gauge, and
\begin{equation}
\Delta_\Omega \equiv \left(\cot \theta \frac{\partial}{\partial\theta} +  \frac{\partial^2}{\partial\theta^2}\right)  +  \left( \frac{1}{\sin^2\theta}\frac{\partial}{\partial \phi}\right) \equiv \Delta_\theta + \Delta_\phi
\end{equation}
as the  angular  Laplacian operator in spherical coordinates. In
harmonic space, Equation (\ref{eq-Delta-lens}) gives a new
contribution to $a_{\ell m}(z)$ (Equation (\ref{eq-almRSD})), which for
simplicity, can be written here using the equality between the $\Psi$ and $\Phi$ fields in linear theory  in case of $\Lambda$CDM:

\begin{equation}
\begin{split}
 a_{\ell m}^\mathrm{lens}(z) &= i^\ell  \ell(\ell +1) \sqrt{\frac{8}{\pi}}
  \int_0^{r(z)} \mathrm{d}r^\prime  \left(\frac{r(z) -r^\prime}{r(z) r^\prime}\right)   \\
 & \times \int \mathrm{d}\mathbf{k}\ Y^\ast_{\ell m} (\hat{\mathbf{k}})\ \Psi(\mathbf{k},z^\prime) j_\ell(kr^\prime) 
\end{split}
\end{equation}

where we have used the property $\Delta_\Omega Y^\ast_{\ell m} =
 -\ell (\ell +1)  Y^\ast_{\ell m}$. Using the result given in
 \citet{2014CQGra..31w4002B}, the transfer function
  $T_\Psi(k,z)$  associated with the $\Psi$ field is related to  $T_D(k,z)$ 
 (Equation (\ref{eq-DV-transfer})) as 
  
\begin{equation}
\begin{split}
\Psi(\mathbf{k},z) &= T_\Psi(k,z) \Psi_\mathrm{in}(\mathbf{k}) \\
T_\Psi(k,z) &= -\frac{3}{2} \Omega_m (1+z) \left( \frac{H_0}{k}\right)^2 T_D(k,z)\,
\end{split}
\end{equation}
 
with $\Omega_m$ as the present  matter density parameter and $H_0$ as the present Hubble parameter. It yields
 
\begin{equation}
\begin{split}
 a_{\ell m}^\mathrm{lens}(z) &= i^\ell  \ell(\ell +1) (-3\Omega_m H_0^2) \sqrt{\frac{2}{\pi}} \\
& \times  \int_0^{z} \frac{\mathrm{d}z^\prime}{H(z^\prime)}  \left(\frac{r(z) -r(z^\prime)}{r(z) r(z^\prime)}\right)  (1+z^\prime)   \\
 & \times \int \frac{\mathrm{d}\mathbf{k}}{k^2}\ Y^\ast_{\ell m} (\hat{\mathbf{k}})\ \Psi_\mathrm{in}(\mathbf{k}) T_D(k,z^\prime) j_\ell(kr(z^\prime)) 
\end{split}
\end{equation}
where we have  changed the integration variable $r^\prime$ to $z^\prime$. 
 
Combining this result with the density and RSD contributions of
Equation (\ref{eq-almRSD}) and the extension given by Equation
(\ref{eq-alm-V}), the expression of $\xi(\theta; z_1, z_2)$ leads to new infinite $\ell$-sums of two types:
\begin{eqnarray}
S_\mathrm{lens-cross} &=& \sum_{\ell=0}^\infty (2\ell +1) P_\ell(\cos\theta) j_\ell^{(n)}(x) \ell (\ell+1)j_\ell(x^\prime)\quad \label{eq-lens-cross}\\
S_\mathrm{lens-lens} &=& \sum_{\ell=0}^\infty (2\ell +1) P_\ell(\cos\theta) \ell^2 (\ell+1)^2 j_\ell(x) j_\ell(x^\prime) \quad \label{eq-lens-lens}.
\end{eqnarray}

The first equation type comes from the cross-correlation of the
lensing term with the density or RSD terms, as well as those concerned
by the $\mathbf{V.n}$ contribution, leading to a $ j_\ell^{(n)}(x)$ factor  ($x=kr(z)$ and similarly $x^\prime = k r(z^\prime)$). The second equation is the autocorrelation of the lensing  term. 

These two equation types can be handled with the following Legendre polynomial property
\begin{equation}
\Delta_\theta P_\ell(\cos \theta) = - \ell (\ell +1) P_\ell(\cos \theta),
\end{equation}
which  yields to the following derivative operator actions on the $A(x, x^\prime,\theta)$ function:
\begin{equation}
\begin{split}
S_\mathrm{lens-cross} &= - \left( \frac{\partial^n   }{\partial x^n}\Delta_\theta \right) A(x, x^\prime,\theta) \\
S_\mathrm{lens-lens}  &= \left( \Delta_\theta^2\right) A(x, x^\prime,\theta),
\end{split}
\end{equation}
where the $\Delta_\theta$ operator acts on $\theta$ in $R(x,x^\prime,
\theta)$ (see Appendix \ref{app-afunc}). With these results we can
include the lensing terms in the computation of $\xi(\theta, z_1,
z_2)$ using the method developed in Section \ref{sec:direct-method}. 

\section{Implementation in \texttt{Angpow}}
\label{sec:angpow}
The formalism we developed is very suitable to implementation within
our public code \texttt{Angpow} \citep{2017arXiv170103592C}, which performs fast and accurate
computations of such highly oscillating integrals. The latest version
now includes the computation of $\xi(\theta, z_1, z_2)$ with density,
RSD, and Doppler terms \footnote{Downloadable from \url{https://gitlab.in2p3.fr/campagne/AngPow}}.

In practice, we compute
\begin{equation}
\bar{\xi}(\theta, z_1, z_2) = \iint \mathrm{d}z\ \mathrm{d}z^\prime W_1(z,z_1,\sigma_1) W_2(z^\prime,z_2,\sigma_2)\ \xi(\theta, z, z^\prime)
\label{eq-xi-thick-shell}
\end{equation}
with $W_i(z,z_i,\sigma_i)$ $(i=1,2)$, a user-defined redshift selection function of typical $\sigma_i$ width centered around $\langle z \rangle = z_i$ (class \texttt{RadSelectBase}). For this release the $\xi(\theta, z, z^\prime)$ function includes the density, RSD contributions developed in Section \ref{sec:new-expansion}, and the  $\mathbf{V.n}$ (Doppler-like) contribution introduced in Section \ref{sec:Vn-terms}. 

\begin{figure}\centering
\includegraphics[width=1.0\linewidth]{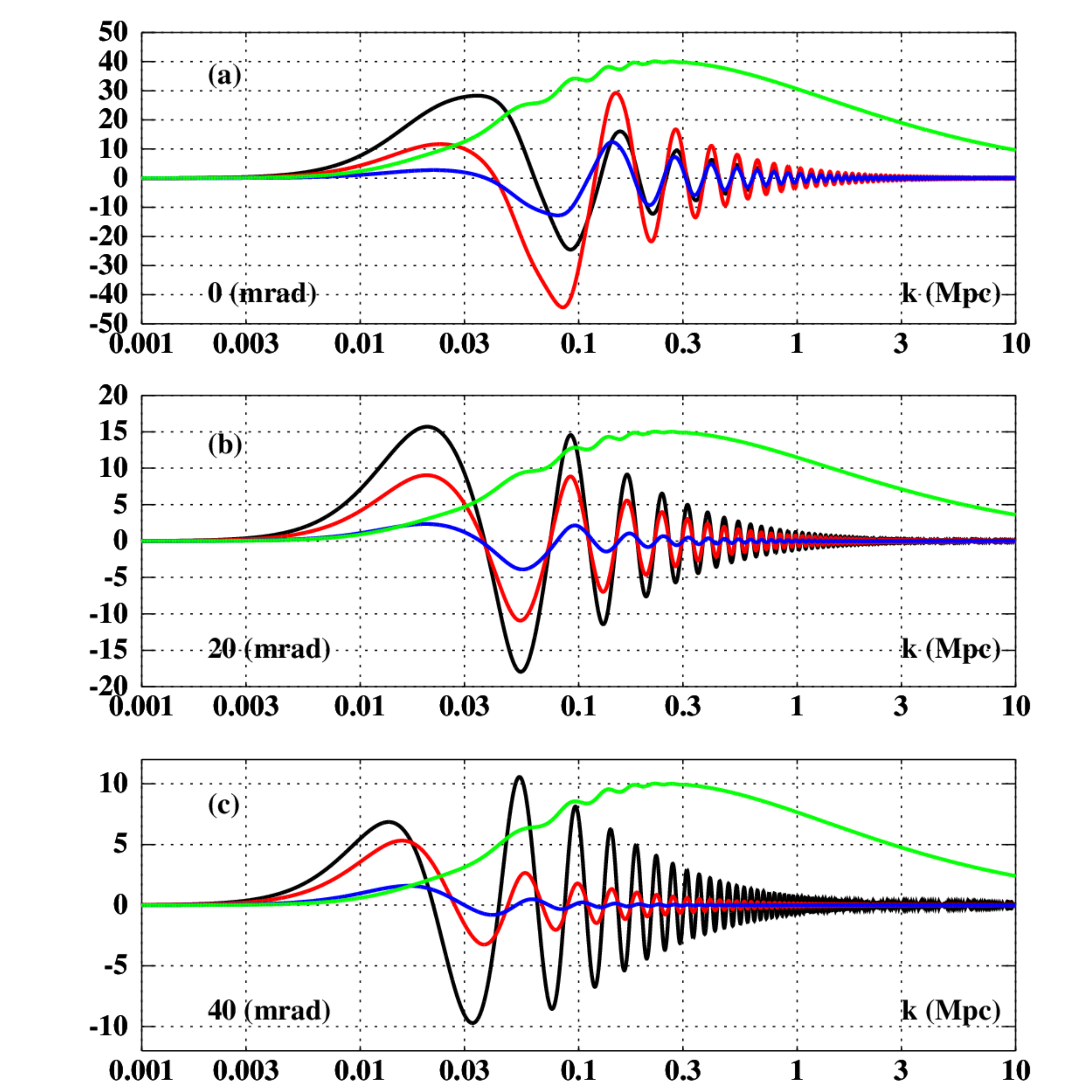}
\caption{Behaviors of the different contributions to the $\xi(\theta, z_1, z_2)$ integral described in Equations (\ref{eq-xi-thick-shell}-\ref{eq-f-3C})   with $z_1=1$, $z_2$ corresponding to $r(z_2)-r(z_1)=50$~Mpc and $\theta =$ 0~mrad in panel (a), 20~mrad in panel (b), and 40~mrad in panel (c). The matter density and  the main RSD contributions and their interference are represented by  black, blue, and red curves, respectively. In each panel,  the $k^2 P|_{z=0}(k)$ function is shown in green in arbitrary units from a standard $\Lambda$CDM cosmology. 
}
\label{fig-afunc}
\end{figure}

Gathering the different contributions, one can write
\begin{equation}
\xi(\theta, z, z^\prime) = \frac{G(z)G(z^\prime)}{2\pi^2}\int \mathrm{d}k\ k^2\ P|_{z=0}(k)\ f(k;\theta, z, z^\prime)
\label{eq:xi-3Calgo}
\end{equation}
with $P|_{z=0}(k)$ determined from the concrete implementation of the
class \texttt{PowerSpecBase} to read an external $(k, P(k))$-tuple
saved by the \texttt{CLASSgal} output. The default growth factor
$G(z)$ is taken from \citet{1991MNRAS.251..128L} and \citet{1992ARA&A..30..499C}, from which we compute the $f_a(z)$ function. The $ f(k;\theta, z,z^\prime)$ implementation reads
\begin{equation}
\begin{split}
f(& k;\theta, z,z^\prime) \\
&  = b b^\prime A - b f_a^\prime \partial^2_{x^{\prime 2}} A  - b^\prime f_a \partial^2_{x^2} A+ f_af_a^\prime \partial^4_{x^2 x^{\prime 2}} A \\
& - b \frac{\tilde{\alpha}^\prime}{x^\prime} \, \partial_{x^\prime} A - b^\prime \frac{\tilde{\alpha}}{x} \, \partial_{x} A + \frac{\tilde{\alpha} f_a^\prime}{x} \partial^3_{x\, x^{\prime 2}} A \\
& + \frac{\tilde{\alpha}^\prime f_a}{x^\prime} \partial^3_{x^2\, x^\prime} A +  \frac{\tilde{\alpha} \tilde{\alpha}^\prime }{x\, x^\prime} \partial^2_{x \, x^\prime} A
\end{split}
\label{eq-f-3C}
\end{equation}
where we have explicitly exhibited the bias $b=b(z)$ factor, $f_a =
f_a(z)$, and $A$ stands for $A(\theta, x, x^\prime)$ with $x = k
r(z)$.  We have contracted the notation of the $A$-function
derivatives as $\partial^{n}_{x^p\, x^{\prime q}}A = \partial^n
A/(\partial x^p \partial x^{\prime q})$, with $p+q=n$.  The
$\tilde{\alpha}(z)$ function is defined from $\alpha(z)$ (Equation (\ref{eq-Vn-contrib})) as $\alpha(z) {\cal H}(z) f_a(z)$. The sub-leading term of the RSD leads to  $\tilde{\alpha}(z) = 2 f_a(z)$ \citep{2012PhRvD..86f3503M}. The prime stands for the $z^\prime$ dependence of the different terms. In Equation (\ref{eq-f-3C}) note the density contribution autocorrelation at the first line, the density-RSD cross-correlation and RSD autocorrelation at the second line, the density-Doppler cross-correlation at the third line, the RSD-Doppler cross-correlation at the fourth line, and the Doppler autocorrelation at the last line. As an illustration, the Figure \ref{fig-afunc} shows the behaviors of the different pieces  entering Equations (\ref{eq-xi-thick-shell}-\ref{eq-f-3C}), taking into account the density and main RSD terms, and Dirac selection functions at $z_1=1$ and $z_2$ such that $r(z_2)-r(z_1)=50$~Mpc, and different $\theta$ values: 0~mrad in panel (a), 20~mrad in panel (b), and 40~mrad in panel (c).

The $\xi(\theta, z, z^\prime)$ function can be computed thanks to the 3C-algorithm developed for \texttt{Angpow} and described in \citet{2017arXiv170103592C}. In brief, this algorithm proceeds as follows:
\begin{enumerate}
\item the total integration $k$ interval (e.g., $[k_\mathrm{min}, k_\mathrm{max}]$) in Equation (\ref{eq:xi-3Calgo}) is cut on several $k$-sub-intervals;
\item  on each sub-interval the functions $$f_1(k) =  G(z)G(z^\prime)/(2\pi^2)\times k^2\ P|_{z=0}(k)$$ and schematically $$f_2(k) = f(k;\theta, z,z^\prime)$$ are projected onto Chebyshev series of order $2^N$;
\item the product of the two Chebyshev series is performed with a $2^{2N}$ Chebyshev series; 
\item then, the integral on the sub-interval is computed thanks to the Clenshaw-Curtis quadrature.   
\end{enumerate}
All the Chebyshev expansions and the Clenshaw-Curtis quadrature are
performed via the DCT-I fast transform of FFTW.

\section{Numerical results}
\label{sec-results}
%
%
%
\begin{figure}\centering
\includegraphics[width=0.8\linewidth]{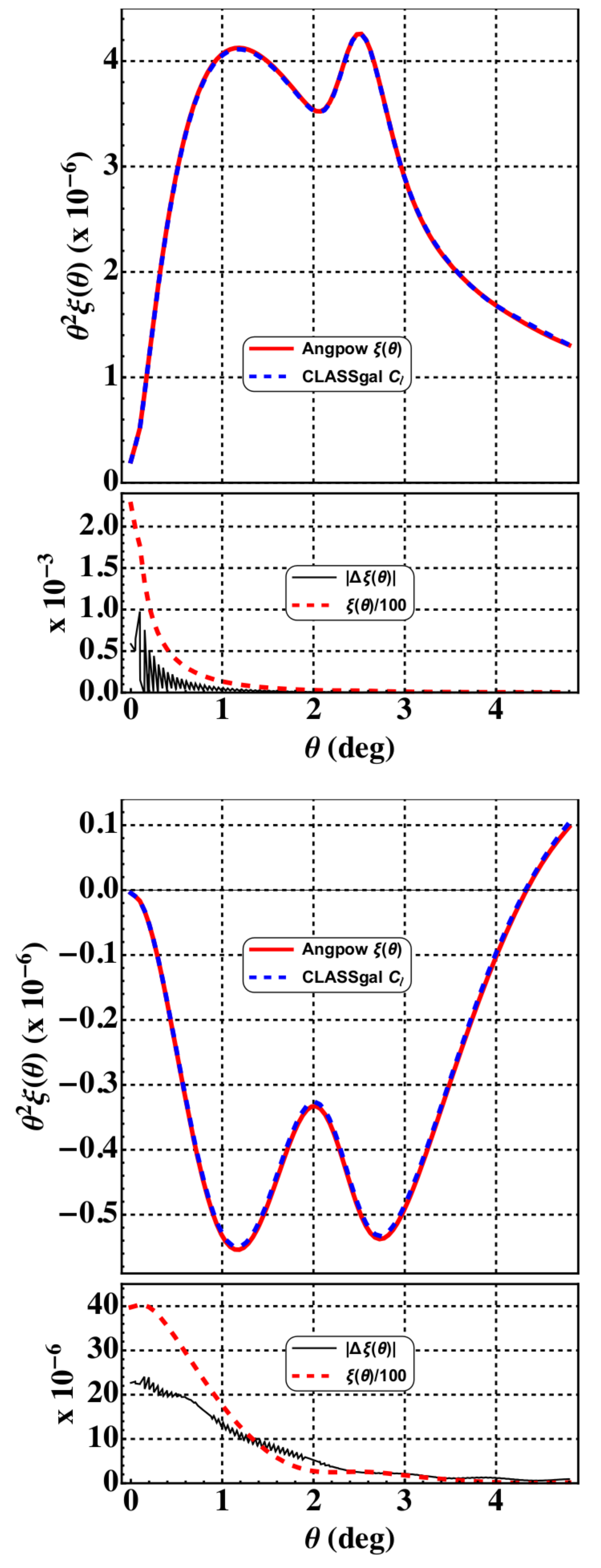}
\caption{Top panel: the $\theta^2 \xi(\theta, z_1, z_2)$ computations with $z_1=z_2=1.00$ and a Gaussian selection function of width $\sigma =0.01$, with redshift-space distortions included. The red curve is the result of the present paper (Equations (\ref{eq-Ctheta-Afunc}) and (\ref{eq:xi-3Calgo})), while the  blue  dashed curves displays the \texttt{CLASSgal} result given by the $C_\ell$ summation. Shown in the bottom panel is the absolute differences $\Delta \xi$ between the present paper and the \texttt{CLASSgal} results (black curve). As a matter of scale, we also display the $\xi(\theta, z_1, z_2)$ spectrum downscaled by a factor 100. Bottom panel(s): with the same color conventions and code parameters we display the results of the cross-correlation between two shells centered at $z_1=1.00$ and $z_2=1.05$ ($\sigma =0.01$).}
\label{fig-auto-cross-angcor-angpow-class}
\end{figure}

To test and benchmark our $\xi$ closed form implementation, we have
chosen to compare its output to the flexible, publicly and well established available software that
performs similar computations, \texttt{CLASSgal} \citep{ClassGal}. Note that the \texttt{CAMBsource}
\citep{2011PhRvD..84d3516C} code is in principle able to perform the same computation.
We have chosen a standard cosmology (for instance we set $h=0.679$, $\Omega_{\mrm{b}} =  0.0483$, $\Omega_{\mrm{m}} =  0.2582$ and $\Omega_{\mrm{k}} = \Omega_{\mrm{fld}} = 0$). We compute with each code the
density+RSD correlation functions in and between two shells at redshifts of
$z_1=1$ and $z_2=1.05$, and a common Gaussian redshift selection width of $\sigma = 0.01$.
For a proper comparison we turned off the Limber approximations in
\texttt{CLASSgal}. We set a cut on the $k$ integral at $k_\texttt{max} = 1~\mathrm{Mpc}^{-1}$
because of memory overload with \texttt{CLASSgal} for larger values. 
In contrast our implementation is not memory-limited.
Since the \texttt{CLASSgal} code only computes the $C_\ell(z_1,z_2)$ power
spectra, afterward we perform the Equation~(\ref{eq-Ctheta-z1z2})
transform. 
We emphasize that in our formalism the $\ell$ sum in Equation
(\ref{eq-abra}) is performed formally up to
infinity, so for comparison with a $C_\ell$-based approach we need to
go to very high multipoles. Fortunately, any $k_\mathtt{max}$ cut that
anyhow always exists in real survey analysis also limits the power in harmonic
space to roughly 
$\ell_\mathtt{max}\simeq \langle r(z)\rangle\, k_\mathtt{max}$, where $\langle r(z)\rangle$ is the mean
comobile distance to the shells. Since in our benchmark setup $\langle r(z)\rangle \simeq 3300$~Mpc, we compute multipoles up to
$\ell_\mathtt{max}=3500$ and no extra apodizing smoothing is necessary to wash out the $\ell_\mathtt{max}$ cut effect, since the $C_\ell$ spectrum decreases quickly enough to 0 at $\ell = \ell_\mathtt{max}$.

A comparison between both computations of the correlation functions is shown in
Figure~\ref{fig-auto-cross-angcor-angpow-class}. 
They agree at the percent level. The main difference is in the CPU
time and Random Access Memory (RAM) used. While \texttt{CLASSgal} runs on 8 threads
lasting around $3$~min and requires 50GB of RAM, our
computation is performed in 30~s using 500~MB of RAM. On 16
threads, the wall time with \texttt{CLASSgal} is essentially unchanged
but requires 100~GB of RAM, while our implementation runs in 18~s with 1~GB of RAM.

\begin{figure*}\centering
\includegraphics[width=0.8\linewidth]{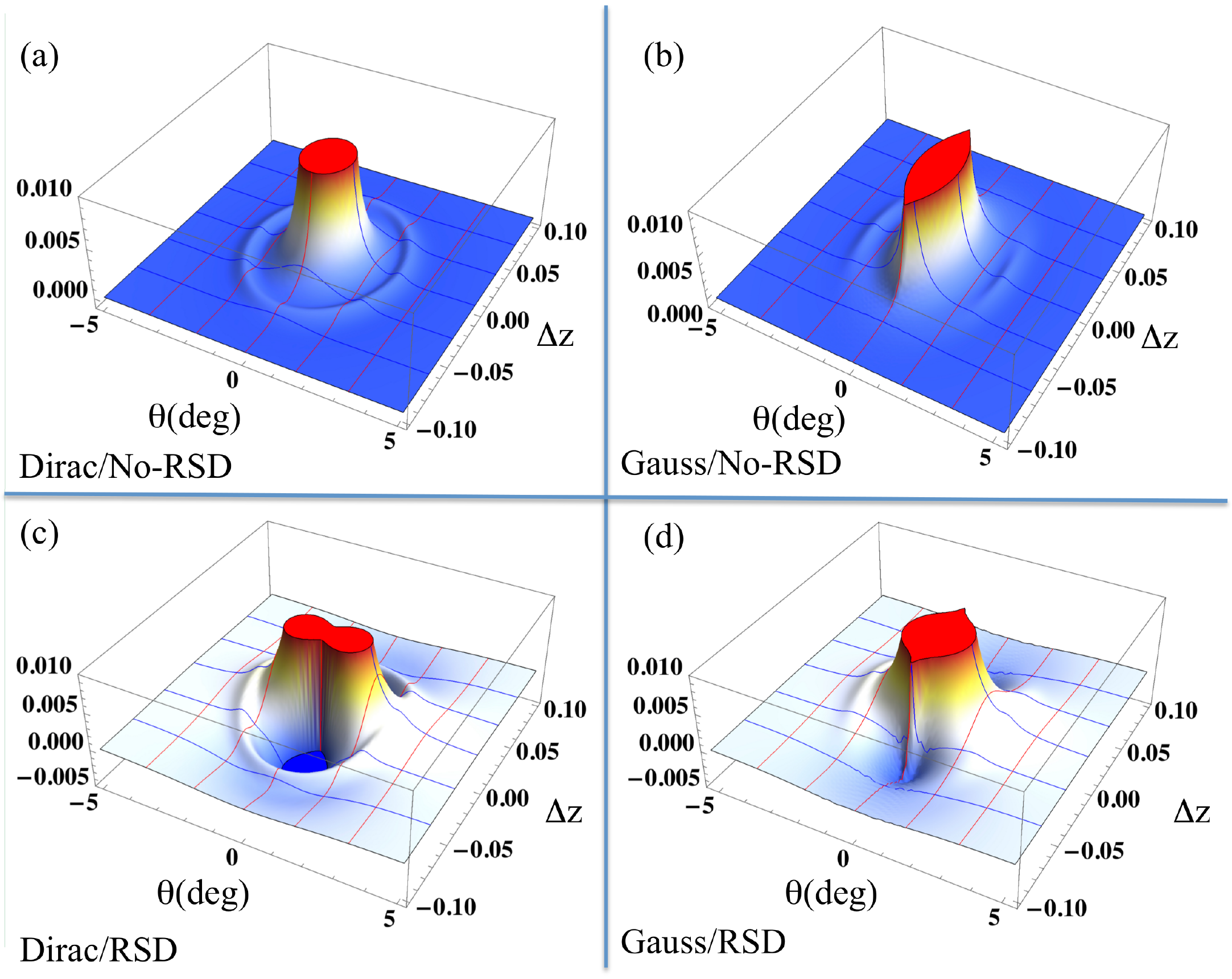}
\caption{Values of $\xi(\theta, z_1, z_2)$ presented as  surface plots
  depending on $\theta$ and $\Delta z=z_2-z_1$ with $z_1=1$, using Dirac (a), (c) or Gaussian
  (b), (d) redshift selection functions and without (a), (b) or with (c),
  (d) RSD terms. For Gaussian selection we have used $\sigma=0.01$ and a
  $5\sigma$ cut. The color scheme is just a one-to-one correspondence of the vertical scale, but it allows one to appreciate the relief of the surface.} 
\label{fig-dirac-gauss-rsd-nosrsd}
\end{figure*}

With such performances we can also display the full
anisotropic $\xi(\theta,z_1, z_2)$ function in the observable space ($\theta$, $\Delta z= z_2-z_1$). With this purpose we fix $z_1$ to 1 and vary $\theta$ and $z_2$. This is similar to the figures displayed in reference \citet{2004ApJ...615..573M}, but we can show the effect of the redshift selection function.
In the upper left plot of Figure \ref{fig-dirac-gauss-rsd-nosrsd},
obtained without RSD and Dirac radial selection functions (panel
(a)), one can discern the isotropic BAO wiggle corresponding to a comoving sound horizon  scale of about 150~Mpc. 
Switching to a Gaussian redshift selection function ($\sigma=0.01$,
panel (b)) stretches out the central peak along the $\Delta z$ axis
(note that no angular resolution has been taken into account) and the BAO
wiggle is washed out along $\theta \approx 0$, while it is preserved at
higher angles for close $z$-shells ($\Delta z \approx 0)$. 
Turning on the RSD contributions, i.e., both the density-RSD cross-correlation and the RSD self-correlation (panel (c) for Dirac
selection and panel (d) for Gaussian selection), tends to shrink the
central correlation peak along the  $\Delta z$ axis and to also
develop a negative correlation for $\theta \approx 0$ and $\Delta z =
\pm 0.05$ corresponding approximately to where the BAO peak sits in
the  Dirac-w/o RSD case (panel (a)). 
The effect of the Gaussian redshift selection function in the presence of
RSD is identical that in the Dirac case. 

 %
\section{Summary and Outlooks}
\label{sec-summary}
We have described a straightforward method to obtain a closed form for the angular correlation function of galaxy counts $\xi(\theta, z_1, z_2)$ (Equation (\ref{eq-Ctheta-Afunc})). The main ingredient to derive it was a reformulation of the problem in harmonic space and an addition theorem (Equation (\ref{eq-abra})), which lead to the simple analytic functions presented in Appendix \ref{app-afunc}. The closed form determined with our approach has been checked to correspond to a previous one after identifying the different expansion coefficients.

The 3C-algorithm detailed in reference \citet{2017arXiv170103592C} and our compact form of $\xi(\theta, z_1, z_2)$  offer a fast and accurate way to compute such correlation functions.
We have implemented, tested, and benchmarked it within our \texttt{Angpow} public code. 
We have presented and discussed the full angular correlation functions in 2D space 
 involving the density-density correlation as well as the RSD-RSD and the RSD-density terms, and compare the accuracy and speed to the public code \texttt{CLASSgal}.
 
Our approach can also include other terms from the relativistic linear perturbation theory, such as the Doppler effect, which implies some simple extra derivatives, or the lensing effects, which requires the angular derivatives of the Laplacian operator.

Our code is fast enough to transform each model into configuration space and to allow us to perform some clean 
parameter inference as if using a Monte-Carlo Markov Chain procedure.
Since it works directly in the observable space, observers are not
forced any more to inject some fiducial cosmology into their large-scale structure data analyses to convert angles and redshift into
distances. Data could then be independent of cosmological assumptions
and be freely used to test the $\Lambda$CDM-General Relativity paradigm, which has been assumed 
in the derivation of the $\xi(\theta, z_1, z_2)$ close form expression of this work.
\acknowledgements
We thank R. Ansari for fruitful discussions throughout the production of this work. 
%
%
\bibliographystyle{aasjournal}
\bibliography{Ctheta}
\appendix
\section{$a_{m_1 m_2}$ and $b_{m_1 m_2}$ coefficients}
\label{sec:ab-coeff}
Here are the expressions of the $a_{m_1 m_2}$ and $b_{m_1 m_2}$
coefficients used in Equation (\ref{eq:xi-tripolar-expansion}) from
reference \citet{2012PhRvD..86f3503M}, where we have replaced $\beta_i$ by $ f_a(z_i)$ and $r$ by $R$ according to our definitions: 
\bea
a_{00}&=&\bra{1+\frac{1}{3}\bra{f_a(z_1)+f_a(z_2)}+\frac{2}{15}f_a(z_1)f_a(z_2)} \zeta_0^2(R)
- \bra{\frac{1}{6}\bra{f_a(z_1)+f_a(z_2)}+\frac{2}{21}f_a(z_1)f_a(z_2)} \zeta_2^2(R)
+ \frac{3}{140}f_a(z_1)f_a(z_2) \;\zeta_4^2(R) \;, \nonumber \\
a_{20}&=& -\bra{\frac{1}{2}f_a(z_1)+\frac{3}{14}f_a(z_1)f_a(z_2)} \zeta_2^2(R) + \frac{1}{28}f_a(z_1)f_a(z_2) \;\zeta_4^2(R) \;, \nonumber \\
a_{02}&=& -\bra{\frac{1}{2}f_a(z_2)+\frac{3}{14}f_a(z_1)f_a(z_2)} \zeta_2^2(R) + \frac{1}{28}f_a(z_1)f_a(z_2) \;\zeta_4^2(R) \;, \nonumber \\
a_{22}&=& \frac{1}{15}f_a(z_1)f_a(z_2) \;\zeta_0^2(R) - \frac{1}{21}f_a(z_1)f_a(z_2) \;\zeta_2^2(R) + \frac{19}{140}f_a(z_1)f_a(z_2) \;\zeta_4^2(R) \;, \nonumber \\
b_{22}&=& \frac{1}{15}f_a(z_1)f_a(z_2) \;\zeta_0^2(R) - \frac{1}{21}f_a(z_1)f_a(z_2) \;\zeta_2^2(R) - \frac{4}{35}f_a(z_1)f_a(z_2) \;\zeta_4^2(R) \;, \nonumber \\
\label{eq:coord_coeff}
\eea

\section{Details on $R(x_1,x_2,\theta)$  and $A(x_1,x_2,\theta)$ function derivatives}
\label{app-afunc}
In this Appendix we provide the detailed expressions for the derivative functions involved in Equation
 (\ref{eq-Ctheta-Afunc}) and also first-order derivatives that can be used in other use cases. To stabilize the oscillation behavior of these functions, it is convenient to perform the derivation with respect to the square of the $R$ function. In the following, we define  $x \equiv x_2-x_1$ and $T\equiv \theta/2$.
We start with the derivatives of the $R^2$ function, which yield
\begin{equation}
\frac{\partial R^2}{\partial x_1} = -2 x+ 4 x_2 \sin^2 T, \qquad
\frac{\partial R^2}{\partial x_2} =  2 x+ 4 x_1 \sin^2 T, \qquad
\frac{\partial^2 R^2}{\partial x_1\partial x_2} = -2+4\sin^2 T, \qquad 
\frac{\partial^2 R^2}{\partial x_1^2} = \frac{\partial^2 R^2}{\partial x_2^2} = 2
\end{equation}
The higher-order derivatives with respect to $x_1$ or $x_2$ are null.
Considering the derivatives of the $R$ function itself yields
\begin{align}
\frac{\partial R}{\partial x_1} &= \frac{1}{2 R}\ \frac{\partial R^2}{\partial x_1}&
\frac{\partial R}{\partial x_2} &= \frac{1}{2 R}\ \frac{\partial R^2}{\partial x_2} \nonumber \\
\frac{\partial^2 R}{\partial x_1^2} &= \frac{1}{4 R^{3/2}}\left[
- \left( \frac{\partial R^2}{\partial x_1} \right)^2 + 4 R^2
\right]&
\frac{\partial^2 R}{\partial x_2^2} &= \frac{1}{4 R^{3/2}}\left[
- \left( \frac{\partial R^2}{\partial x_2} \right)^2 + 4 R^2
\right] \nonumber \\
\frac{\partial^2 R}{\partial x_1\partial x_2} &= \frac{1}{4 R^{3/2}}\left[
- \frac{\partial R^2}{\partial x_1} \frac{\partial R^2}{\partial x_2} + 2 R^2 \frac{\partial^2 R^2}{\partial x_1\partial x_2} 
\right]&
\frac{\partial^3 R}{\partial x_1^2\partial x_2} &= \frac{1}{8 R^{5/2}}\left[
\frac{\partial R^2}{\partial x_2}  \left(
3 \left(\frac{\partial R^2}{\partial x_1}\right)^2 - 4 R^2
\right) 
- 4 R^2 \frac{\partial R^2}{\partial x_1} \frac{\partial^2 R^2}{\partial x_1\partial x_2} 
\right]\nonumber \\
\frac{\partial^3 R}{\partial x_1\partial x_2^2} &= \frac{1}{8 R^{5/2}}\left[
\frac{\partial R^2}{\partial x_1}  \left(
3 \left(\frac{\partial R^2}{\partial x_2}\right)^2 - 4 R^2
\right) 
- 4 R^2 \frac{\partial R^2}{\partial x_2} \frac{\partial^2 R^2}{\partial x_1\partial x_2} 
\right]& 
\frac{\partial^3 R}{\partial x_1\partial x_2^2} &= \frac{1}{8 R^{5/2}}\left[
\frac{\partial R^2}{\partial x_1}  \left(
3 \left(\frac{\partial R^2}{\partial x_2}\right)^2 - 4 R^2
\right) 
- 4 R^2 \frac{\partial R^2}{\partial x_2} \frac{\partial^2 R^2}{\partial x_1\partial x_2} 
\right] \nonumber \\
\frac{\partial^4 R}{\partial x_1^2\partial x_2^2} &=\frac{1}{16 R^{7/2}}\left[
-15 \left(\frac{\partial R^2}{\partial x_1}\right)^2  \left(\frac{\partial R^2}{\partial x_2}\right)^2 
+12 R^2 \left\{
\left( \frac{\partial R^2}{\partial x_1} \right)^2 \right. \right.&& \nonumber \\
&\left. \left. + \left( \frac{\partial R^2}{\partial x_2} \right)^2 
+2 \frac{\partial R^2}{\partial x_1} \frac{\partial R^2}{\partial x_2} \frac{\partial^2 R^2}{\partial x_1\partial x_2} 
\right\}
- 8 R^4 \left\{
2 + \left( \frac{\partial^2 R^2}{\partial x_1\partial x_2} \right)^2 
\right\}
\right]&&
\end{align}
For the sake of completeness, the derivatives of $\sinc(x)=j_0(x)$ using $\cos(x)$ and $\sinc(x)$ functions read
\begin{align}
\sinc^{(1)}(x) &= \frac{1}{x}\left(\cos x - \ \sinc\ x \right),&
\sinc^{(2)}(x) &=  -\frac{1}{x^2}\left(2 \cos x + (x^2 -2) \ \sinc\ x \right) \nonumber \\
\sinc^{(3)}(x) &=  \frac{1}{x^3}\left( -(x^2-6) \cos x + 3(x^2 -2) \ \sinc\ x \right),&
\sinc^{(4)}(x) &=  \frac{1}{x^4}\left( 4(x^2-6) \cos x + (x^4-12x^2 +24) \ \sinc\ x \right)
\end{align}
Note that the Taylor expansions at $x=0$ of these derivative functions are
\begin{align}
\sinc^{(1)}(x) &\approx -\frac{1}{3}x +\frac{1}{30}x^3 - \frac{1}{840}x^5 +\dots,& 
\sinc^{(2)}(x) &\approx -\frac{1}{3} + \frac{1}{10}x^2 - \frac{1}{168}x^4 + \dots \nonumber \\
\sinc^{(3)}(x) &\approx \frac{1}{5}x -\frac{1}{42}x^3 + \frac{1}{1080}x^5 + \dots,& 
\sinc^{(4)}(x) &\approx \frac{1}{5}- \frac{1}{14}x^2 + \frac{1}{216}x^4 + \dots
\end{align}
Using the above results, the expressions of the $A$-function derivatives used in Equation (\ref{eq-Ctheta-Afunc}) and also used in case to add $j^\prime(x)$ contributions to Equation (\ref{eq-Doppler-term}), are 
\begin{align}
\frac{\partial A}{\partial x_1} &= \sinc^{(1)}(R)  \frac{\partial R}{\partial x_1},&
\frac{\partial A}{\partial x_2} &= \left. \frac{\partial A}{\partial x_1} \right|_{x_1 \rightarrow x_2}\nonumber \\
\frac{\partial^2 A}{\partial x_1^2} &= \sinc^{(2)}(R)\left(\frac{\partial R}{\partial x_1} \right)^2 +\sinc^{(1)}(R)\frac{\partial^2 R}{\partial x_1^2},&
\frac{\partial^2 A}{\partial x_2^2}& = \left.\frac{\partial^2 A}{\partial x_1^2}\right|_{x_1 \rightarrow x_2}\nonumber \\
\frac{\partial^2 A}{\partial x_1\partial x_2} &= \sinc^{(2)}(R) \frac{\partial R}{\partial x_1} \frac{\partial R}{\partial x_2} + \sinc^{(1)}(R)  \frac{\partial^2 R}{\partial x_1\partial x_2},&
\frac{\partial^3 A}{\partial x_1\partial x_2^2} &= \frac{\partial R}{\partial x_1} \left\{
\sinc^{(3)}(R)  \left( \frac{\partial R}{\partial x_2} \right)^2 
+ \sinc^{(2)}(R)  \frac{\partial^2 R}{\partial x_2^2}
\right\}\nonumber \\
&&&+ 2\ \sinc^{(2)}(R) \frac{\partial R}{\partial x_2} \frac{\partial^2 R}{\partial x_1\partial x_2}
+  \sinc^{(2)}(R) \frac{\partial^3 R}{\partial x_1\partial x_2^2} \nonumber \\
\frac{\partial^3 A}{\partial x_1^2\partial x_2} &= \left. \frac{\partial^3 A}{\partial x_1\partial x_2^2} \right|_{x_1 \rightarrow x_2},&
\frac{\partial^4 A}{\partial x_1^2\partial x_2^2} &= \sinc^{(4)}(R) \left(\frac{\partial R}{\partial x_1} \right)^2 \left(\frac{\partial R}{\partial x_2} \right)^2 \nonumber \\
&&& + \sinc^{(3)}(R) \left\{ 
\left( \frac{\partial R}{\partial x_1}\right)^2 \frac{\partial^2 R}{\partial x_2^2} 
+ 4 \frac{\partial R}{\partial x_1} \frac{\partial R}{\partial x_2}  \frac{\partial^2 R}{\partial x_1\partial x_2} 
+ \left( \frac{\partial R}{\partial x_2}\right)^2 \frac{\partial^2 R}{\partial x_1^2} 
\right\} \nonumber \\
&&&+  \sinc^{(2)}(R) \left\{
2  \left( \frac{\partial^2 R}{\partial x_1\partial x_2} \right)^2 + 2 \frac{\partial R}{\partial x_1} \frac{\partial^3 R}{\partial x_1\partial x_2^2}
+  \frac{\partial^2 R}{\partial x_1^2}  \frac{\partial^2 R}{\partial x_2^2} \right. \nonumber \\
&&&\left. + 2 \frac{\partial R}{\partial x_2} \frac{\partial^3 R}{\partial x_1^2\partial x_2}\right\} +  \sinc^{(1)}(R) \frac{\partial^4 R}{\partial x_1^2\partial x_2^2}
\label{eq-A-derivatives}
\end{align}
The $\Delta_\theta$ operator action on the $A$-function involved in the lensing term (Section \ref{sec:extansions}) reads
\begin{equation}
\begin{split}
\Delta_\theta(A(x_1,x_2,\theta)) &=\left(\cot\theta \frac{\partial R}{\partial \theta} +\frac{\partial^2 R}{\partial \theta^2}\right) \sinc^{(1)}(R) + \left( \frac{\partial R}{\partial \theta}\right)^2   \sinc^{(2)}(R) \\
\Delta_\theta^2(A(x_1,x_2,\theta)) &= \left(\cot\theta \csc^2\theta \frac{\partial R}{\partial \theta} - (1+\csc^2\theta)\frac{\partial^2 R}{\partial \theta^2} + 2 \cot\theta \frac{\partial^3 R}{\partial \theta^3} + \frac{\partial^4 R}{\partial \theta^4}\right) \sinc^{(1)}(R) \\
&+ \left( -(1+\csc^2\theta)\left( \frac{\partial R}{\partial \theta}\right)^2 + 3\left( \frac{\partial^2 R}{\partial \theta^2}\right)^2 + \frac{\partial R}{\partial \theta}\left[
6 \cot\theta \frac{\partial^2 R}{\partial \theta^2} + 4 \frac{\partial^3 R}{\partial \theta^3}\right]
\right) \sinc^{(2)}(R) \\
& + 2 \left( \cot\theta \frac{\partial R}{\partial \theta} + 3 \frac{\partial^2 R}{\partial \theta^2}
\right)\left(\frac{\partial R}{\partial \theta}\right)^2  \sinc^{(3)}(R) \\
&+ \left(\frac{\partial R}{\partial \theta}\right)^4 \sinc^{(4)}(R)
\end{split}
\end{equation}
with the derivatives of the $R(x_1,x_2,\theta)$ function with respect to $\theta$
\begin{equation}
\begin{split}
R \frac{\partial R}{\partial \theta} &= 4 x_1 x_2 \cos\theta \sin\theta\\
R^3\frac{\partial^2 R}{\partial \theta^2} &= 2x_1x_2\left( 2(x_1+x_2)^2\cos2\theta - x_1 x_2(3+\cos4\theta) \right)\\
R^5 \frac{\partial^3 R}{\partial \theta^3} &= - 4 x_1 x_2 \sin2\theta\left( 
 2(x_1^4+x_2^4) + 8x_1x_2 (x_1^2+x_2^2) + 7 x_1^2 x_2^2 + x_1x_2(x_1 x_2 \cos4\theta -2 \cos 2 \theta\, (x_1+x_2)^2)
\right)\\
R^7\frac{\partial^4 R}{\partial \theta^4}  &=  -2 x_1 x_2 \left(4 \left(x_1+x_2\right){}^2 \left(2 x_1^4+8 x_1^3 x_2+11 x_1^2 x_2^2+8 x_1 x_2^3+2 x_2^4\right) \cos 2 \theta \right. \\
&+ \left. x_1 x_2 \left(4
\left(x_1^4+4 x_1^3 x_2-5 x_1^2 x_2^2+4 x_1 x_2^3+x_2^4\right)\cos 4 \theta -7 \left(4 x_1^4+16 x_1^3 x_2+13 x_1^2 x_2^2+16 x_1 x_2^3+4 x_2^4\right) \right.\right. \\
&+ \left. \left. x_1
x_2 \left(- x_1 x_2 \cos 8 \theta +4  \left(x_1+x_2\right){}^2\cos 6 \theta  \right)\right)\right)   
\end{split}
\end{equation}
\section{Equality between the tripolar and harmonic result}
\label{app:equality}
We demonstrate here that both approaches in Sections \ref{sec:tripolar} and
\ref{sec:new-expansion} lead to equivalent results, although in
different forms. We take Equation (\ref{eq-Ctheta-Afunc})  as the initial expansion and
identify the coefficients proportional to $f_a(z_1)$ ($f_a(z_2))$ and
$f_a(z_1)f_a(z_2)$. Note that the first term of expression
(\ref{eq-Ctheta-Afunc}) is easily  identified as the $a_{00}$ one in the
expression (\ref{eq:xi-tripolar-expansion}) just simply because $A(x_1,x_2,\theta) =j_0(k R)$, as a result of Equation (\ref{eq-abra}) (note that $x_i = k r_i$). Thus, using a symbolic algebra package 
 we can show that  the term proportional to $f_a(z_1)$ can be expanded as
\begin{equation}
\frac{\partial^2 A}{\partial x_1^2} = - \frac{1}{3	} j_0(k R) + \frac{1}{6} j_2(k R) + \frac{1}{2}  j_2(kR) \cos 2\phi_1
\end{equation}
One identifies the corresponding factors of $a_{00}$ and $a_{20}$ thanks to the minus sign of Equation (\ref{eq-Ctheta-Afunc}), as the $\zeta_\ell^2$ expansion can be interpreted as $j_\ell$ expansion. By the same method one can identify at which $a_{00}$ and $a_{20}$ factors correspond to $\partial^2 A/\partial x_2^2$.

To get the corresponding $j_\ell$ expansion of the $\partial^4
A/\partial x_1^2\partial x_2^2$ term (\ref{eq-A-derivatives}) we use the following identities:
\onecolumngrid
\begin{equation}
\begin{split}
\left(\frac{\partial R}{\partial x_1} \right)^2 \left(\frac{\partial R}{\partial x_2} \right)^2 &=
\frac{1}{4}\left(\vphantom{\frac{1}{4}}1+\cos 2\phi_1\right)\left(\vphantom{\frac{1}{4}}1+\cos 2\phi_2\right) \\
(kR)\left\{ \left( \frac{\partial R}{\partial x_1}\right)^2 \frac{\partial^2 R}{\partial x_2^2} 
+ 4 \frac{\partial R}{\partial x_1} \frac{\partial R}{\partial x_2}  \frac{\partial^2 R}{\partial x_1\partial x_2} 
+ \left( \frac{\partial R}{\partial x_2}\right)^2 \frac{\partial^2 R}{\partial x_1^2} \right\}
&= \frac{1}{2} \left(\vphantom{\frac{1}{2}}
1 - \cos 2\phi_1\cos 2\phi_2 + 2 \sin 2\phi_1 \sin 2\phi_2 \right) \\
(kR)^2\left\{  2  \left( \frac{\partial^2 R}{\partial x_1\partial x_2} \right)^2 + 2 \frac{\partial R}{\partial x_1} \frac{\partial^3 R}{\partial x_1\partial x_2^2}
+  \frac{\partial^2 R}{\partial x_1^2}  \frac{\partial^2 R}{\partial x_2^2} + 2 \frac{\partial R}{\partial x_2}  \frac{\partial^3 R}{\partial x_1^2\partial x_2}\right\} &=
-\frac{1}{4} \left(\vphantom{\frac{1}{4}}
1 + 3 (\cos 2\phi_1 + \cos 2\phi_2) -7\cos 2\phi_1 \cos 2\phi_2 \right. \\
& \phantom{-\frac{1}{4} \left(1 + 3 (\cos 2\phi_1 + \cos 2\phi_2) - \right. }\left. \vphantom{\frac{1}{4}} + 8\sin 2\phi_1 \sin 2\phi_2 \right) \\
&= - (kR)^3 \frac{\partial^4 R}{\partial x_1^2\partial x_2^2} 
\end{split}
\label{eq-Atophi}
\end{equation}
while the derivatives of the $\sinc(x)$ can be expanded as a linear combination of $j_\ell(x)$ with $\ell=0,2,4$ according to:
\begin{equation}
\begin{split}
\sinc^{(4)}(x) &= \frac{1}{5} j_0(x) - \frac{4}{7} j_2(x) + \frac{8}{35} j_4(x) \\
\frac{\sinc^{(3)}(x)}{x} &= \frac{1}{5} j_0(x) + \frac{1}{7} j_2(x) - \frac{2}{35} j_4(x) \\
-\frac{\sinc^{(1)}(x)}{x^3} + \frac{\sinc^{(2)}(x)}{x^2} &=  
\frac{1}{15}j_0(x) + \frac{2}{21}j_2(x) + \frac{1}{35}j_4(x). 
\end{split}
\label{eq-sinc2jl}
\end{equation}
Then, gathering the different $j_\ell$ factors of the combination of Equations (\ref{eq-Atophi}) and (\ref{eq-sinc2jl}), we recover the contributions to $a_{00}$, $a_{20}$, $a_{02}$, $a_{22}$, and $b_{22}$. 
So, we find that the Equations (\ref{eq:xi-tripolar-expansion}) and (\ref{eq-Ctheta-Afunc}) match perfectly, as expected. 
\end{document}